\documentclass{webofc}

\usepackage[varg]{txfonts}   
\usepackage{hyperref}
\usepackage{url}
\hypersetup{colorlinks=true,citecolor=blue,urlcolor=blue,linkcolor=blue}
\begin{document}
\title{
Finite density lattice QCD without extrapolations }

\author{
\firstname{Alexander} \lastname{Adam}\inst{1}
\and
\firstname{Szabolcs} \lastname{Bors\'anyi}\inst{1}
\and
\firstname{Zolt\'an} \lastname{Fodor}\inst{2,3,1,4,5}
\and
\firstname{Jana N.} \lastname{Guenther}\inst{1}
\and
\firstname{Paolo} \lastname{Parotto}\inst{6}
\and
\firstname{Attila} \lastname{P\'asztor}\inst{4,7}
\and
\firstname{Ludovica} \lastname{Pirelli}\inst{1}\fnsep\thanks{\email{pirelli@uni-wuppertal.de}}     
\and
\firstname{Chik Him} \lastname{Wong}\inst{1}
}

\institute{Department of Physics, Wuppertal University, Gaussstr.  20, D-42119, Wuppertal, Germany
\and
Department of Physics, Pennsylvania State University,
State College, PA 16801, USA 
\and
Institute for Computational and Data Sciences, Pennsylvania State University,
University Park, PA 16802, USA
\and
Institute  for Theoretical Physics, ELTE E\"otv\"os Lor\' and University, 
P\'azm\'any P. s\'et\'any 1/A, H-1117 Budapest, Hungary
\and
J\"ulich Supercomputing Centre, Forschungszentrum J\"ulich, D-52425 J\"ulich, Germany
\and
Dipartimento di Fisica, Universit\`a di Torino and INFN Torino, 
Via P. Giuria 1, I-10125 Torino, Italy
\and
MTA-ELTE Lend\"ulet "Momentum"  Strongly Interacting Matter Research Group, Budapest, Hungary
          }

\abstract{Finite density lattice QCD usually relies on extrapolations in baryon chemical potential ($\mu_B$), be it Taylor expansion, T' expansion (\cite{Borsanyi:2021sxv}) or analytical continuation. However, their range of validity is difficult to control. In the canonical formulation, the baryon density is the parameter of the system, not $\mu_B$.
Here we demonstrate that we can access finite density QCD in the canonical formulation with physical quark masses. 
We present first results with both the strangeness ($n_S$) and baryon ($n_B$) densities as parameters. Specifically, we compute the QCD pressure and chemical potentials as functions of $n_B$ and $n_S$. 
Our computations rely on
high-statistics simulations with 2+1 4HEX-staggered fermions. 

}

\maketitle

\section{Introduction}
\label{intro}
\begin{figure*}
\centering
\vspace*{1cm}  
\includegraphics[width=6cm,clip]{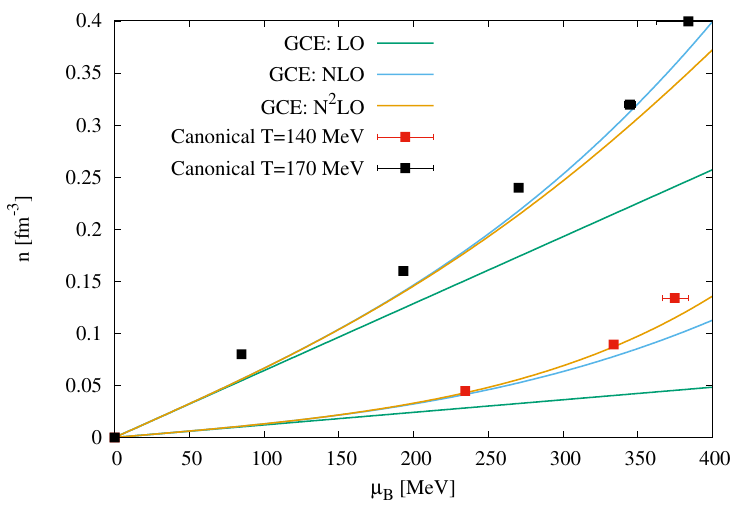}
\includegraphics[width=6cm,clip]{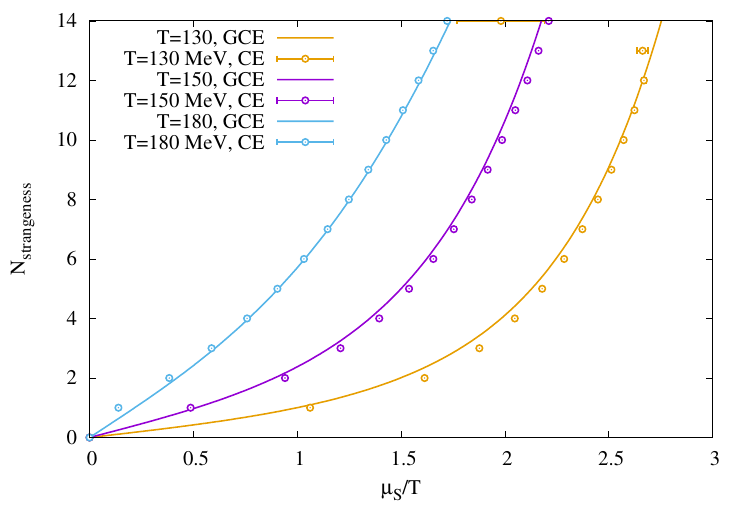}
\caption{Left plot:  baryon particle density $n=N/V$ in terms of the baryochemical potential $\mu_B$ in the grand canonical ensemble and in the canonical ensemble for $T=140,\,170 $ MeV, corresponding respectively to the simulation volumes $V\sim22, \ 13 \, \rm{fm}^3$. $n$ is computed in the GCE with a Taylor expansion. $\mu_B$ is computed in  the CE as a finite difference.
Analogously we compute $N_S$ as a function of $\mu_S$ (right plot). For one flavour only it is easier to compute a higher number of Fourier coefficients.}
\label{mu_vs_n}       
\end{figure*}

The canonical formulation of QCD at finite density plays an important role in the phenomenological modelling of heavy-ion collisions. 
Writing the partition function in the canonical ensemble (CE) is interesting in lattice QCD (LQCD) also for theoretical reasons. The partition function in the grand canonical ensemble (GCE) $Z_{GC}(T,V,\mu_B)$ can be simulated with LQCD methods only at zero or imaginary $\mu_B$ because of the sign problem; analytical continuation, Taylor or reweighting methods are then used to reach the real $\mu_B$ region. With the first two methods an extrapolation in $\mu_B$ is needed and we cannot escape a systematic error. Reweighting from $\mu_B=0$  does not imply any extrapolation, but it is affected by an overlap problem. If staggered fermions are used in the lattice action, there is also an additional problem caused by rooting, that gives rise to non-analyticities in the free energy ~\cite{Golterman:2006rw,Borsanyi:2023tdp}. Canonical results at fixed number of particles $N$ have by definition no systematics in $\mu_B$.
\\
The canonical partition function $Z_C(T,V,N)$ can be obtained from the grand canonical partition function $Z_{GC}(T,V,\mu_B)$ via a Fourier transform ~\cite{Alexandru:2005ix,deForcrand:2006ec}:
\begin{equation}
\begin{array}{ccc}
\label{fourier_transf}
    Z_C(T,V,N) = \frac{1}{2\pi}\int_0^{2\pi}  Z_{GC}(T,V,\mu_B) \, \cos(N \phi_B) \, d\phi_B & \, & \mu_B=i \phi_B T. \end{array}
\end{equation}
We compute the grand canonical partition function $Z_{GC}(T,V,\mu_B)$ via reweighting from zero and imaginary values of $\mu_B$.
 The baryochemical potential is now a derived observable and it is defined as a discrete derivative of the canonical  thermodynamical potential $F(T,V,N)$:
\begin{equation}
\label{mu_eq}
\begin{array}{ccc}
    F(T,V,N) = - T \log Z(T,V,N) & \, & \mu_B(T,V,N) = F(T,V,N)-F(T,V,N-1)  \end{array}
\end{equation}
We compare the GCE and CE relation between the density  and baryochemical potential in figure ~\ref{mu_vs_n}.

\section{Results}
\label{Results}
In the following we show results computed with 2+1 flavors at physical quark masses, on a $16^3\times8$ lattice. We used a 4HEX staggered fermionic action in the simulations (~\cite{Capitani:2006ni,Borsanyi:2023wno}).
\subsection{Volume effects in the CE}
\label{vol_effects}
\begin{figure*}
\centering
\vspace*{1cm}       
\includegraphics[width=6cm,clip]{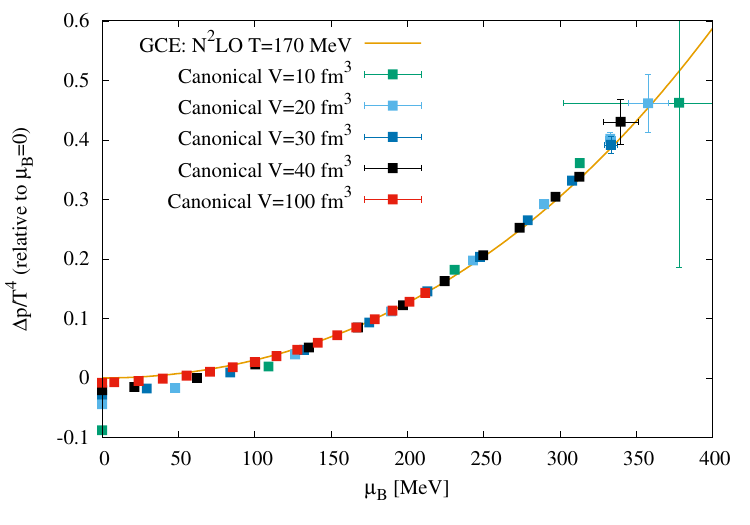}
\includegraphics[width=6cm,clip]{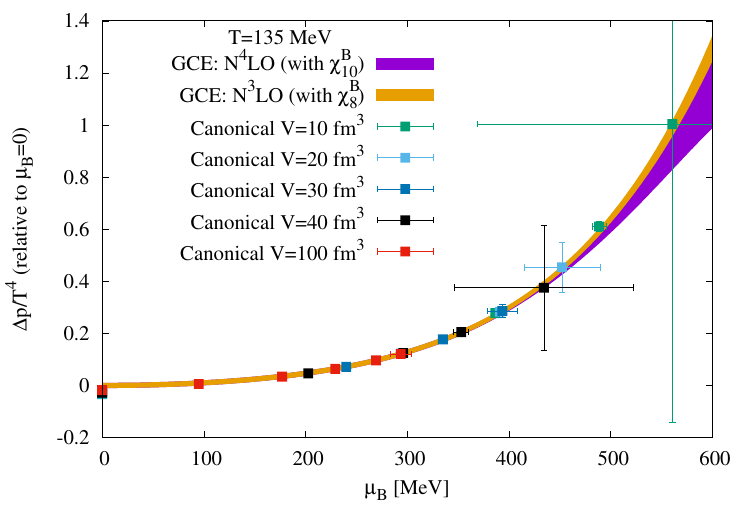}
\caption{$\Delta p/T^4$ as a function of $\mu_B$ for several volumes $V$, see eq.(\ref{pressure}). The simulation volume $V_0$ is $\sim 13 \rm{fm}^3$ for the left plot ($T=170$ MeV) and $\sim 25 \rm{fm}^3$ for the right plot ($T=135$ MeV). GCE lines and CE datapoints tend to agree at higher volumes.}
\label{voldep}      
\end{figure*}

The pressure is the volume derivative of the canonical free energy. In our plots we show its difference with respect to the grand canonical pressure at $\mu_B=0$:
\begin{eqnarray}
\label{pressure}
\frac{\Delta p}{T^4}=
    \frac{p_C(T,V,N)}{T^4} 
    -\frac{p_{GC}(\mu_B=0)}{T^4} 
    &=& \frac{1}{T^3} \frac{1}{Z_C(T,V,N)} \frac{\partial Z_C(T,V,N)}{\partial V}\nonumber\\
    &=& \frac{1}{VT^3}
    \frac{\frac{1}{\pi} \int_0^\pi Z_{GC}(i\phi_B) \log\left(\frac{Z_{GC}^{V/V_0}(i\phi_B)}{Z_{GC}(0)}\right) \cos(N\phi) d\phi_B}%
{\frac{1}{\pi} \int_0^\pi Z_{GC}^{V/V_0}(i\phi_B) \cos(N\phi_B) d\phi}. \nonumber\\
\end{eqnarray}
$Z_{GC}$ is computed at one fixed simulation volume $V_0$, so  we introduced a volume rescaling factor $V/V_0$ in $Z_{GC}$  to make the derivative. That is also useful to compute $\Delta p/T^4$ at different volumes other than the simulated one, as in figure ~\ref{voldep}.  Canonical and grand canonical results agree only in the thermodynamic limit. 
\subsection{Comparison with HRG}
\label{hrg}
\begin{figure*}
\centering
\vspace*{1cm}       
\includegraphics[width=6cm,clip]{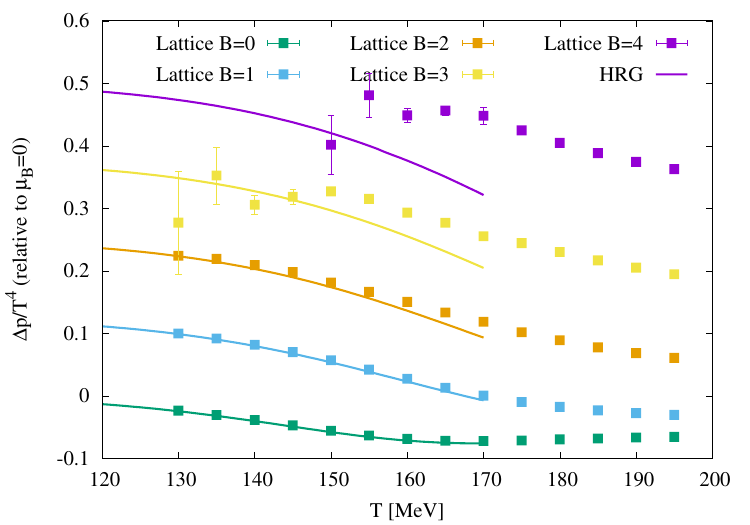}
\includegraphics[width=6cm,clip]{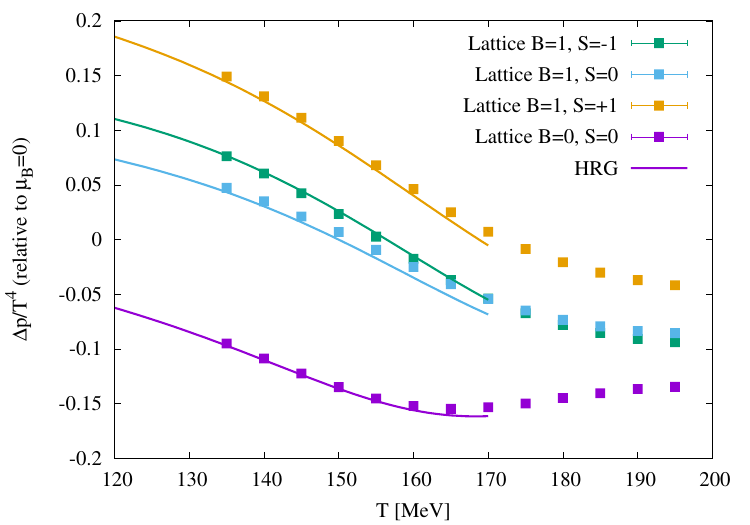}
\caption{Comparison with HRG. Left plot: $\Delta p/T^4$ at fixed $B=0,...,4$. Right plot: $\Delta p/T^4$ at $B=1$, $S=0,1,-1$. Also the case $B=0,\, S=0$ is shown. }
\label{hrg_comp}       
\end{figure*}

In figure ~\ref{hrg_comp} we look at the comparison with Hadron Resonance Gas results in the temperature range $T=[120,200] $ MeV ~\cite{Vovchenko:2019pjl}. 
In the left plot we show $\Delta p/T^4$ at different fixed baryon numbers $B=0,...,4$. As expected, for higher $B$ the results agree with the non-interacting case only for small temperatures. \\
All these results were actually computed in a mixed setup, canonical in $B$ but still grand canonical in $\phi_S$, following the previous references on the topic ~\cite{Alexandru:2005ix,deForcrand:2006ecsz_}. We now make one step forward and we compute $Z_C$ both in baryon $B$ and strangeness $S$ numbers with a two dimensional Fourier transform:
\begin{equation}
\label{ZBS}
Z_{BS}=\frac{1}{(2\pi)^2} \int_{0}^{2\pi} d\phi_{S} \int_{0}^{2\pi} d\phi_{B} \, Z_{GC}(\phi_{B},\phi_{S})\,e^{-i\phi_{S}S}e^{-i\phi_{B}B}.
\end{equation}
The electric charge $Q$ is still out of the picture. 
We then compute $\Delta p/T^4$ at fixed $B, S$ and we compare them with HRG in the right plot of figure ~\ref{hrg_comp}.

\subsection{Conclusions}
A canonical formulation of lattice QCD is employed at physical quark masses on a $16^3 \times 8$ lattice volume. We have results both in a mixed approach (canonical in $B$, grand canonical in $\phi_S$) as well as in a $B, S$ formulation. The electric charge $Q$ is still not taken into consideration. Although the errors on $\Delta p/T^4$ are still big at higher $B$, they have the advantage to be only statistical, since canonical results are by definition not affected by sistematic errors due to $\mu_B$ extrapolations. 

\section{Acknowledgments}
This work is also supported by the MKW NRW under the funding code NW21-024-A. 
The authors gratefully acknowledge the Gauss Centre for Supercomputing e.V. (\url{www.gauss-centre.eu}) for funding this project by providing computing time on the GCS Supercomputer Juwels-Booster at Juelich Supercomputer Centre. We acknowledge the EuroHPC Joint Undertaking for awarding this project access to the EuroHPC supercomputer LUMI, hosted by CSC (Finland) and the LUMI consortium through a EuroHPC Extreme Access call.

\bibliography{biblio}

\end{document}